\documentstyle[12pt,epsf]{article}

\begin{document}
\topskip 2cm 
\begin{titlepage}

\hspace*{\fill}\parbox[t]{4cm}{EDINBURGH 97/26 \\ MSUHEP-71111
\\ Nov 12, 1997}

\vspace{2cm}

\begin{center}
{\large\bf Virtual Next-to-Leading Corrections to the Impact Factors
in the High-Energy Limit} \\
\vspace{1.5cm}
{\large Vittorio Del Duca} \\
\vspace{.5cm}
{\sl Particle Physics Theory Group,\,
Dept. of Physics and Astronomy\\ University of Edinburgh,\,
Edinburgh EH9 3JZ, Scotland, UK}\\
\vspace{.5cm}
and\\
\vspace{.5cm}
{\large Carl R. Schmidt} \\
\vspace{.5cm}
{\sl Department of Physics and Astronomy\\
Michigan State University\\
East Lansing, MI 48824, USA}\\
\vspace{1.5cm}
\vfil

\begin{abstract}

We compute the virtual next-to-leading corrections to the impact factors
or off-shell coefficient functions in the high-energy limit.
When combined with the known real corrections, these 
results will provide the complete NLO corrections to the impact
factors, which are necessary to use the BFKL resummation at NLL for jet
production at both lepton-hadron and hadron-hadron colliders.

\end{abstract}
\end{center}

\end{titlepage}

\section{Introduction}
\label{sec:0}

Semi-hard strong-interaction processes, which are characterized by two large
and disparate kinematic scales, typically lead to cross sections 
containing large logarithms.  Two examples of this type of process 
are Deep Inelastic Scattering (DIS) at small $x$ and hadronic dijet production 
at large rapidity intervals $\Delta y$.  In DIS the logarithm that 
appears is $\ln(1/x)$, with $x \simeq Q^2/s$ 
the squared ratio of the momentum transfer to the photon-hadron 
center-of-mass energy.  In large-rapidity dijet production the large 
logarithm is $\Delta y\simeq\ln(\hat s/|\hat t|)$, with
$\hat s$ the squared parton center-of-mass energy and $|\hat t|$ of the
order of the squared jet transverse energy.  These logarithms will 
arise in a perturbative calculation at each order in the coupling 
constant $\alpha_{s}$.  Alternatively, if the logarithms are large 
enough, it is preferable to include them through an all-order resummation 
in the leading logarithmic (LL) approximation performed by means of the
Balitsky-Fadin-Kuraev-Lipatov (BFKL) equation \cite{FKL}-\cite{bal}.

In order to show how the latter comes about we consider dijet production at 
large rapidity intervals as a paradigm process. At the lowest order,
${\cal O}(\alpha_s^2)$, the underlying parton process is dominated
in the $\hat s \gg |\hat t|$ limit
by gluon exchange in the $t$-channel.  Thus the functional form of 
the amplitudes for gluon-gluon, gluon-quark or quark-quark scattering is the
same; they differ only by the color strength in the parton-production
vertices.  We can then write the cross section for 
dijet production in the high-energy limit in the following factorized 
form \cite{DS}
\begin{equation}
{d\sigma\over d^2 p_{a'\perp} d^2 p_{b'\perp} dy_{a'} dy_{b'}}\, =\,
x_A^0 f_{eff}(x_A^0,\mu_F^2)\, x_B^0 f_{eff}(x_B^0,\mu_F^2)\,
{d\hat\sigma_{gg}\over d^2 p_{a'\perp} d^2 p_{b'\perp}}\, ,\label{mrfac}
\end{equation}
where $a'$ and $b'$ label the forward and backward outgoing jet, 
respectively, and the effective $pdf$'s are 
\begin{equation}
f_{eff}(x,\mu_F^2) = G(x,\mu_F^2) + {4\over 9}\sum_f
\left[Q_f(x,\mu_F^2) + \bar Q_f(x,\mu_F^2)\right], \label{effec}
\end{equation}
where the sum is over the quark flavors.  The cross section for
gluon-gluon scattering at leading order in the high-energy limit is 
\begin{equation}
{d\hat\sigma_{gg}^{0}\over d^2 p_{a'\perp} d^2 p_{b'\perp}}\ =\
\biggl[{C_A\alpha_s\over p_{a'\perp}^2}\biggr] \,
{1\over2}\delta^{2}(p_{a'\perp}+p_{b'\perp}) \,
\biggl[{C_A\alpha_s\over p_{b'\perp}^2}\biggr] \ ,
\label{cross0}
\end{equation}
where the Casimir is $C_{A}=N_{c}=3$.

At higher orders, powers of $\ln(\hat s/|\hat t|)$ will arise, which can 
be resummed to all orders in $\alpha_{s}\ln(\hat s/|\hat t|)$, {\it 
i.e.} to LL accuracy, by the BFKL equation.
The factorization formula (\ref{mrfac}) is 
unchanged, and the only modification is in the gluon-gluon scattering 
cross section which becomes \cite{mn}
\begin{equation}
{d\hat\sigma_{gg}^{0}\over d^2 p_{a'\perp} d^2 p_{b'\perp}}\ =\
\biggl[{C_A\alpha_s\over p_{a'\perp}^2}\biggr] \,
f(q_{a\perp},q_{b\perp},\Delta y) \,
\biggl[{C_A\alpha_s\over p_{b'\perp}^2}\biggr] \ ,
\label{cross}
\end{equation}
with $\Delta y = y_{a'}-y_{b'}$ and $q_{i\perp}$ the momenta transferred in 
the $t$-channel, {\it i.e.} $q_{a\perp}=p_{a'\perp}$ and 
$q_{b\perp}=-p_{b'\perp}$.  The function $f(q_{a\perp},q_{b\perp},
\Delta y)$ is the
Green's function associated with the gluon exchanged in the 
$t$-channel.  It is process independent and given in the LL 
approximation by the solution of the
BFKL equation.  This equation is a two-dimensional
integral equation which describes the evolution in transverse momentum
 of the gluon propagator exchanged in the $t$-channel.  
If we transform to moment space via
\begin{equation}
f(q_{a\perp},q_{b\perp},\Delta y) \ =\ \int {d\omega\over 2\pi i}\, 
e^{\omega\Delta y}\, 
f_{\omega}(q_{a\perp},q_{b\perp})\label{moment}
\end{equation}
we can write the BFKL equation as
\begin{equation}
\omega\, f_{\omega}(q_{a\perp},q_{b\perp})\, =
{1\over 2}\,\delta^2(q_{a\perp}-q_{b\perp})\, +\, {\alpha_s N_c\over \pi^2} 
\int {d^2k_{\perp}\over k_{\perp}^2}\,
K(q_{a\perp},q_{b\perp},k_{\perp})\, ,\label{bfklb}
\end{equation}
where the kernel $K$ is given by
\begin{equation}
K(q_{a\perp},q_{b\perp},k_{\perp}) = f_{\omega}(q_{a\perp}+k_{\perp},
q_{b\perp}) - {q_{a\perp}^2\over k_{\perp}^2 + 
(q_{a\perp}+k_{\perp})^2}\, f_{\omega}(q_{a\perp},q_{b\perp})
\, .\label{kern}
\end{equation}
The first term in the kernel accounts for the emission of a real gluon of 
transverse momentum $k$ and the second term accounts 
for the virtual radiative corrections.  
Eq.~(\ref{bfklb}) has been derived in the
{\sl multi-Regge kinematics}, which presumes that the produced gluons are
strongly ordered in rapidity and have comparable transverse momenta 
\begin{equation}
y_{a'} \gg y \gg y_{b'}; \qquad |p_{a'\perp}|\simeq|k_\perp|
\simeq|p_{b'\perp}|\, .\label{mrk}
\end{equation}
The solution to the BFKL equation is
\begin{equation}
f(q_{a\perp},q_{b\perp},\Delta y)\ =\ {1\over (2\pi)^2 
q_{a\perp} q_{b\perp}} 
\sum_{n=-\infty}^{\infty} e^{in\tilde\phi}\, \int_{-\infty}^{\infty} d\nu\, 
e^{\omega(\nu,n)\Delta y}\, \left(q_{a\perp}^2\over q_{b\perp}^2
\right)^{i\nu}\, ,\label{solc}
\end{equation}
where $\tilde\phi$ is the azimuthal angle
between $q_{a\perp}$ and $q_{b\perp}$
and $\omega(\nu,n)$ is the eigenvalue of
the BFKL equation whose maximum $\omega(0,0)=4\ln{2}N_c\alpha_s/\pi$
yields the known power-like growth of $f$ in energy~\cite{fkl2,bal}.

The vertices in square brackets in eq.~(\ref{cross}) account for the
scattering of an off-shell and an on-shell gluon to produce a gluon
$g^*\, g \rightarrow g$, and
are characteristic of the scattering process under consideration.
In the literature they have been called impact factors or LO 
off-shell coefficient functions.  They are known also for photon-photon
scattering \cite{bal}, with impact factor $\gamma g^* \rightarrow q\, 
\bar q$; for heavy-quark photoproduction, with impact factor
$\gamma g^* \rightarrow Q\, \bar Q$ \cite{er,cch}; for
leptoproduction, with impact factor $\gamma^* g^* \rightarrow Q\, 
\bar Q$ \cite{cch}; for hadroproduction, with impact factors
$g g^* \rightarrow Q\, \bar Q$ \cite{er} and $g^*\, g^* \rightarrow 
Q\, \bar Q$ \cite{cch,ce}; for hadroproduction via electroweak boson 
exchange, with impact factors $V^* g^* \rightarrow Q_i\, \bar Q_j$,
with $V= W, Z$ \cite{cc}; for direct photoproduction
in hadron-hadron scattering, with impact factor 
$q\, g^* \rightarrow q\, \gamma$ \cite{er}; and for DIS at small $x$
and forward-jet production in DIS, with impact factor $\gamma^* g^* 
\rightarrow q\, \bar q$ which may be obtained from
$\gamma^* g^* \rightarrow Q\, \bar Q$ in the massless limit of the
heavy quark \cite{stef}. In all the cases above the parton cross
section is obtained by multiplying the process-independent gluon
propagator (\ref{solc}) with the appropriate impact factors. If we
label by ${\cal F}$ a generic impact factor, then the parton cross section
(\ref{cross}) for a generic process in the high-energy limit is
\begin{equation}
\hat\sigma_{LL} \sim {\cal F}_{LO}(q_{a\perp})\, 
f_{LL}(q_{a\perp},q_{b\perp},\Delta y) \,
{\cal F}_{LO}(q_{b\perp})\, ,\label{gener}
\end{equation}
where the subscripts stress the accuracy to which the gluon propagator
and the impact factors are computed.

The BFKL theory, being a LL resummation and not an exact
calculation, makes a few approximations which, even though formally 
subleading, may be important for any phenomenological purposes:
\begin{itemize}
\item[i)] The BFKL resummation is performed at fixed coupling constant, 
thus any variation in its scale, $\alpha_s(\nu^2)=\alpha_s(\mu^2) - 
b_0\ln(\nu^2/\mu^2)\alpha_s^2(\mu^2) + \dots$, with $b_0= (11N_c-2N_f)/
12\pi$ and $N_f$ the number of quark flavors, would appear in the 
next-to-leading-logarithmic (NLL) terms, because it yields terms
of ${\cal O}(\alpha_s^n\ln(\nu^2/\mu^2)\ln^{n-1}(\hat s/|\hat t|))$.
\item[ii)] From the kinematics of two-parton production at
$\hat s\gg|\hat t|$ we identify the rapidity interval between the tagging
jets as $\Delta y\simeq\ln(\hat s/|\hat t|)\simeq\ln(\hat s/k_{\perp}^2)$,
however, we know from the exact kinematics that $\Delta y =
\ln(\hat s/|\hat t|-1)= \ln(\hat u/\hat t)$
and $|\hat t| = k_{\perp}^2 (1+\exp(-\Delta y))$,
therefore the identification of the rapidity interval $\Delta y$
with $\ln(\hat s/|\hat t|)$ is correct up to terms of $O(\hat t/\hat s)$.
\item[iii)] Because of the strong rapidity ordering (\ref{mrk}), there
are no collinear divergences in the LL resummation in the BFKL
ladder.  Jets are determined
only to leading order and accordingly have no non-trivial structure.
\item[iv)] Finally, energy and longitudinal-momentum are not conserved 
in the LL limit.  Effectively, this means that the momentum fractions
$x_{A(B)}$ of the incoming partons are not evaluated exactly, which may
induce large numerical errors in certain BFKL predictions.
In the exact kinematics, if $n+2$ partons are produced 
along the ladder, we have
\begin{equation}
x_{A(B)} = {p_{a'\perp}\over\sqrt{s}} e^{(-)y_{a'}} +  
\sum_{i=1}^n {k_{i\perp}\over\sqrt{s}} e^{(-)y_i}
+ {p_{b'\perp}\over\sqrt{s}} e^{(-)y_{b'}}\, ,\label{nkin}
\end{equation}
where the minus sign in the exponentials of the right-hand side applies
to the subscript $B$ on the left-hand side. In the BFKL theory, the LL
approximation and the kinematics (\ref{mrk}) imply that
in the determination of $x_A$ ($x_B$) in eq.~(\ref{mrfac}) only the 
first (last) term in eq.~(\ref{nkin}) is kept, 
\begin{eqnarray}
x_A^0 &=& {p_{a'\perp}\over\sqrt{s}} e^{y_{a'}}\, ,\nonumber\\
x_B^0 &=& {p_{b'\perp}\over\sqrt{s}} e^{-y_{b'}}\, .\label{nkin0}
\end{eqnarray}
A comparison within dijet
production of the exact ${\cal O}(\alpha_s^3)$ three-parton production
with the truncation of the BFKL ladder to
${\cal O}(\alpha_s^3)$ shows that the LL approximation may severely
underestimate the exact evaluation of the $x$'s (\ref{nkin}),
and therefore entail sizable violations of energy-longitudinal momentum
conservation \cite{DS2}.
Energy-momentum conservation at each
stage in the gluon emission in the BFKL ladder may
be achieved through a Monte Carlo implementation of the
BFKL equation (\ref{bfklb}) \cite{carl,os}.
However, the weights used to determine the gluon emission
and the virtual radiative corrections in the Monte Carlo
are still fixed by eq.(\ref{kern}), which is computed using multigluon
amplitudes at LL accuracy.
\end{itemize}

In order to improve on all the points highlighted above the
NLL corrections to the BFKL equation need to be calculated. This 
calculation is close to an end as all the real \cite{real}-\cite{ptlipqq}
and virtual \cite{fl}-\cite{ffk} corrections to the relevant vertices
have been computed. However, in a production cross section
the calculation of the process-independent NLL corrections to the
gluon propagator exchanged in the $t$-channel must be matched by
impact factors or off-shell coefficient functions computed at the same
accuracy. Using the notation of eq.~(\ref{gener}),
\begin{eqnarray}
\hat\sigma_{NLL} &\sim& {\cal F}_{NLO}(q_{a\perp})\, f_{LL}(q_{a\perp},
q_{b\perp},\Delta y) \,
{\cal F}_{LO}(q_{b\perp})\nonumber\\ &+& {\cal F}_{LO}(q_{a\perp})\, 
f_{NLL}(q_{a\perp},q_{b\perp},\Delta y) \, {\cal F}_{LO}(q_{b\perp})
\label{nlgener}\\ &+& 
{\cal F}_{LO}(q_{a\perp})\, f_{LL}(q_{a\perp},q_{b\perp},\Delta y) \,
{\cal F}_{NLO}(q_{b\perp})\, .\nonumber
\end{eqnarray}
Thus, for each process of interest the corresponding NLO impact factor
must be computed.

In this paper we compute the 1-loop corrections to the impact factors.
We begin in section 2 by reviewing the LL calculation, as well as the NLO
corrections to the impact factors arising from real parton emissions.
Throughout this paper we work with fixed helicities to organize the results.
In section 3 we turn to the 1-loop virtual corrections to the impact factors.
We obtain them from the known 1-loop $g\,g\rightarrow g\,g$ and $q\, 
q\rightarrow q\, q$ helicity amplitudes, by expanding in the 
high-energy limit.  Our main results are then the 1-loop corrections to 
the $g^*\, g \rightarrow g$ vertex, eqs.~(\ref{vert}) and (\ref{elev}), and 
to the $g^*\, q \rightarrow q$ vertex, eq.~(\ref{lqver}).  These corrections
are given in the
conventional dimensional regularization (CDR) or t'Hooft-Veltman (HV)
schemes and also in the dimensional reduction scheme.  They are 
compared with previous results in the CDR scheme, eqs.~(\ref{fffl}), 
(\ref{twel}), and 
(\ref{ffqvert}), which have been obtained in a different manner.
When combined with the known real ${\cal O}(\alpha_{s})$ corrections, these 
results provide the complete NLO corrections to the impact factors.

\section{Radiative corrections in the high-energy limit}
\label{sec:a}

\subsection{LL corrections to ${\cal O}(\alpha^3_s)$}
\label{sec:aa}

In the high-energy limit\footnote{
For the remainder of this paper we use $s$, $t$, and $u$ without
the hat's for the partonic kinematic variables.}
 $s\gg |t|$, the amplitude for 
$g_a\,g_b\to g_{a'}\,g_{b'}$ scattering, with all external
gluons outgoing, may be written \cite{FKL}, \cite{ptlip}
\begin{equation}
M^{aa'bb'\,{\rm tree}}_{\nu_a\nu_{a'}\nu_{b'}\nu_b} = 2  s
\left[i g\, f^{aa'c}\, C^{gg(0)}_{-\nu_a\nu_{a'}}(-p_a,p_{a'}) \right]
{1\over t} \left[i g\, f^{bb'c}\, C^{gg(0)}_{-\nu_b\nu_{b'}}(-p_b,p_{b'}) 
\right]\, ,\label{elas}
\end{equation}
where the $\nu$'s label the helicities and the vertices $g^*\, g 
\rightarrow g$ are given by
\begin{equation}
C_{-+}^{gg(0)}(-p_a,p_{a'}) = -1 \qquad C_{-+}^{gg(0)}
(-p_b,p_{b'}) = - {p_{b'\perp}^* \over p_{b'\perp}}\, ,\label{centrc}
\end{equation}
with $p_{\perp}=p_x+ip_y$ the complex transverse momentum.
The $C$-vertices transform
into their complex conjugates under helicity reversal,
$C_{\{\nu\}}^*(\{k\}) = C_{\{-\nu\}}(\{k\})$. The helicity-flip
vertex $C^{(0)}_{++}$ is subleading in the high-energy limit.
The square of the amplitude (\ref{elas}), integrated over the phase
space, yields the gluon-gluon production rate to leading order,
$O(\alpha_s^2)$.
For gluon-quark or quark-quark scattering, we only need to exchange
the structure constants with color matrices in the fundamental representation
and change the vertices $C^{gg(0)}$ to $C^{\bar q q(0)}$ \cite{thuile}.

Next, we consider the $O(\alpha_s)$ corrections to this process in the 
high-energy limit.  In order to do that, we
must consider the emission of an additional gluon, i.e. the 
$g_a\,g_b\to g_{a'}\,g\,g_{b'}$ scattering amplitude, 
in the multi-Regge kinematics (\ref{mrk}).
The scattering amplitude is 
\begin{eqnarray}
M^{tree}_{g g \rightarrow g g g} &=& 
2 {s} \left[i g\, f^{aa'c}\, C_{-\nu_a\nu_{a'}}^{gg}(-p_a,p_{a'})
\right]\, {1\over t_a}\, \label{three}\\ & &\quad
\times\  \left[i g\,f^{cdc'}\, 
C^g_{\nu}(q_a,q_b)\right]\, {1\over t_b}\, \left[i g\,
f^{bb'c'}\, C_{-\nu_b\nu_{b'}}^{gg}(-p_b,p_{b'}) \right]\, ,\nonumber
\end{eqnarray}
where $ t_i \simeq - |q_{i\perp}|^2$
and the Lipatov vertex $g^*\, g^* \rightarrow g$ \cite{ptlip,lip}, is
\begin{equation}
C^g_+(q_a,q_b) = \sqrt{2}\, {q^*_{a\perp} q_{b\perp}\over k_{\perp}}\, 
.\label{lipeq}
\end{equation}
The square of the amplitude (\ref{three}), 
integrated over the phase space of the intermediate gluon in
multi-Regge kinematics (\ref{mrk}) yields an 
$O(\alpha_s\ln( s/| t|))$ correction to gluon-gluon scattering.
This real correction, however, is infrared divergent.
To complete the $O(\alpha_s)$ corrections, and to cancel the infrared
divergence, we must compute the 1-loop gluon-gluon amplitude
in the LL approximation.
The virtual radiative corrections to eq.~(\ref{elas}) in
the LL approximation are obtained, to all orders
in $\alpha_s$, by replacing \cite{FKL,zvi}
\begin{equation}
{1\over t} \to {1\over t} 
\left({s\over -t}\right)^{\alpha(t)}\, ,\label{sud}
\end{equation}
in eq.~(\ref{elas}), with $\alpha(t)$ related to the loop 
transverse-momentum integration
\begin{equation}
\alpha(t) \equiv g^2 \alpha^{(1)}(t) = \alpha_s\, N_c\, t \int 
{d^2k_{\perp}\over (2\pi)^2}\, {1\over k_{\perp}^2
(q-k)_{\perp}^2}\qquad t = q^2 \simeq - q_{\perp}^2\, ,\label{allv}
\end{equation}
and $\alpha_s = g^2/4\pi$. 
The infrared divergence in eq.~(\ref{allv}) can be regulated in 4
dimensions with an infrared-cutoff mass. Alternatively, 
using dimensional regularization
in $d=4-2\epsilon$ dimensions, the integral
in eq.~(\ref{allv}) is performed in $2-2\epsilon$ dimensions, yielding
\begin{equation}
\alpha(t) = g^2 \alpha^{(1)}(t) = 2 g^2\, N_c\, {1\over\epsilon} 
\left(\mu^2\over -t\right)^{\epsilon} c_{\Gamma}\, ,\label{alph}
\end{equation}
with
\begin{equation}
c_{\Gamma} = {1\over (4\pi)^{2-\epsilon}}\, {\Gamma(1+\epsilon)\,
\Gamma^2(1-\epsilon)\over \Gamma(1-2\epsilon)}\, .\label{cgam}
\end{equation}
Adding the 1-loop gluon-gluon 
amplitude, multiplied by its tree-level counterpart,
to the square of the amplitude (\ref{three}), 
integrated over the phase space of the intermediate gluon, cancels
the infrared divergences and yields a finite $O(\alpha_s\ln(
s/| t|))$ correction to gluon-gluon scattering.

\subsection{NLL corrections to ${\cal O}(\alpha^3_s)$}
\label{sec:ab}

In order to compute the real $O(\alpha_s)$ corrections to gluon-gluon 
scattering which are not accompanied by a $\ln( s/| t|)$, {\it 
i.e.} that are NLL, 
we must relax the strong rapidity ordering between the produced gluons
(\ref{mrk}).  We must allow for the production of two gluons (and to this
accuracy also of a $q\bar q$ pair) with similar rapidity,
\begin{equation}
\left.\begin{array}{c} y_{a'} \simeq y \gg y_{b'}\\ y_{a'} \gg y 
\simeq y_{b'} \end{array}\right\} \qquad |p_{a'\perp}|\simeq|k_\perp|
\simeq|p_{b'\perp}|\, .\label{qmrk}
\end{equation}
The three-gluon production amplitude for the first rapidity ordering is 
\cite{real,ptlipnl,thuile}
\begin{eqnarray}
\lefteqn{ M^{gg}(-p_a,-\nu_a; p_{a'},\nu_{a'}; k,\nu; p_{b'},\nu_{b'}; 
-p_b,-\nu_b)}
\nonumber\\ & & 
= 2 s \left\{ C^{g\,g\,g}_{-\nu_a\nu_{a'}\nu}
(-p_a,p_{a'},k) \left[ (ig)^2\, f^{aa'c} f^{cdc'} {1\over\sqrt{2}}\,q 
A_{\Sigma\nu_i}(-p_a,p_{a'},k) + \left(\begin{array}{c} p_{a'}
\leftrightarrow k\\ a'\leftrightarrow d \end{array}\right) \right] 
\right\} \nonumber\\ & &\quad \times {1\over t}\, \left[ig\, f^{bb'c'}
C^{g\,g}_{-\nu_b\nu_{b'}}(-p_b,p_{b'}) \right]\, ,\label{nllfg}
\end{eqnarray}
where we have enclosed the production vertex $g^*\, g \rightarrow g\, g$
of gluons $p_{a'}$ and $k$ in curly brackets, and with 
$\sum\nu_i=-\nu_a+\nu_{a'}+\nu$ and
\begin{eqnarray}
C^{g\,g\,g}_{-++}(-p_a,p_{a'},k) = 1\, ; & & C^{g\,g\,g}_{+-+}(-p_a,p_{a'},k) 
= {1 \over\left(1+{k^+\over p_{a'}^+}\right)^2}\, ; \nonumber\\ 
C^{g\,g\,g}_{++-}(-p_a,p_{a'},k) = {1 \over\left(1+{p_{a'}^+\over k^+} 
\right)^2}\, ; & & A_+(-p_a,p_{a'},k) = 2\, {p_{b'\perp}\over p_{a'\perp}}
{1\over k_{\perp} - p_{a'\perp} {k^+\over p_{a'}^+}}\, .\label{treposb}
\end{eqnarray}
The vertex 
$C^{g\,g\,g}_{+++}(-p_a,p_{a'},k)$ is subleading to the required accuracy.

The amplitude for the production of a $q\bar q$ pair, $g\, g
\rightarrow \bar q q g$, for the first rapidity
ordering of eq.~(\ref{qmrk}) is \cite{ptlipqq,thuile},
\begin{eqnarray}
\lefteqn{ M^{\bar{q}q}(-p_a,-\nu_a; p_{a'},\nu_{a'}; k,-\nu_{a'}; 
p_{b'},\nu_{b'}; 
-p_b,-\nu_b)} \nonumber\\ & & = 2 s \left\{\sqrt{2}\, g^2\, 
C^{g\,\bar{q}\,q}_{-\nu_a\nu_{a'}-\nu_{a'}}(-p_a,p_{a'},k)
\left[\left(\lambda^{c'} \lambda^a\right)_{d\bar{a'}}
A_{-\nu_a}(p_{a'},k) + \left(\lambda^a \lambda^{c'}\right)_{d\bar{a'}}
A_{-\nu_a}(k,p_{a'}) \right] \right\} 
\nonumber\\ & &\quad \times {1\over  t}\, \left[ig\, f^{bb'c'} 
C^{g\,g}_{-\nu_b\nu_{b'}}(-p_b,p_{b'}) \right]\, ,\label{forwqq}
\end{eqnarray}
with $p_{a'}$ the antiquark, and the vertex $g^*\, g \rightarrow 
\bar q q$ in curly brackets, with $A$ defined in eq.(\ref{treposb})
and $C^{g\,\bar{q}\,q}$ given by,
\begin{eqnarray}
C^{g\,\bar{q}\,q}_{++-}(-p_a,p_{a'},k) &=& {1 \over 2} \sqrt{p_{a'}^+\over k^+}
{1 \over\left(1+{p_{a'}^+\over k^+} \right)^2} \nonumber\\
C^{g\,\bar{q}\,q}_{+-+}(-p_a,p_{a'},k) &=& {1 \over 2} \sqrt{k^+\over p_{a'}^+}
{1 \over\left(1+{k^+\over p_{a'}^+} \right)^2}\, .\label{cqqa}
\end{eqnarray}
The scattering amplitude $q\, g \rightarrow q g g$, from which the
vertex $g^*\, q \rightarrow g\, q$ is extracted, may be found in 
ref.~\cite{thuile}.

The square of the amplitude (\ref{nllfg}), integrated over the 
phase space (\ref{qmrk}) yields an $O(\alpha_s)$ correction
to gluon-gluon scattering.  It is, however, infrared 
divergent, since the vertex $A$ in eq.(\ref{treposb})
has a collinear divergence as $2k\cdot p_{a'}
\rightarrow 0$, and a soft divergence as $k\rightarrow 0$.
The square of the amplitude (\ref{forwqq}), integrated over the 
phase space (\ref{qmrk}) yields an $O(\alpha_s)$ correction
to gluon-gluon scattering, which is only collinearly divergent.
In this case the soft divergence of the vertex $A$ in eq.(\ref{treposb})
is eliminated by the vanishing of the $C$-vertices (\ref{cqqa}) 
as $k\rightarrow 0$, in accordance with the soft quark limit.

We next consider the virtual radiative corrections to the gluon-gluon 
amplitude.  In order to go beyond the LL approximation 
we need a prescription that
allows us to disentangle the virtual corrections to the vertices 
(\ref{centrc}) from the ones that reggeize the gluon (\ref{sud}).
Such a prescription is supplied by
the general form of the high-energy scattering amplitude, 
arising from a reggeized gluon in the adjoint representation of 
$SU(N_c)$ passed
in the $t$-channel\footnote{Other color structures do occur in the
high-energy limit, but they do not contribute at NLL.  We show this
explicitly for the absorptive part of the 1-loop amplitude in appendix
C.}.
In the helicity basis of eq.~(\ref{elas}) this is given by \cite{fl}
\begin{eqnarray}
M^{aa'bb'}_{\nu_a\nu_{a'}\nu_{b'}\nu_b} &=& s
\left[i g\, f^{aa'c}\, C^{gg}_{-\nu_a\nu_{a'}}(-p_a,p_{a'}) \right]
{1\over t} \left[\left({s\over -t}\right)^{\alpha(t)} +
\left({-s\over -t}\right)^{\alpha(t)} \right] \nonumber\\ &&\times
\left[i g\, f^{bb'c}\, C^{gg}_{-\nu_b\nu_{b'}}(-p_b,p_{b'}) 
\right]\, ,\label{elasb}
\end{eqnarray}
where now 
\begin{equation}
\alpha(t) = g^2 \alpha^{(1)}(t) + g^4 \alpha^{(2)}(t) + O(g^6)\,
,\label{alphb}
\end{equation}
and
\begin{equation}
C^{gg} = C^{gg(0)} + g^2 C^{gg(1)} + O(g^4)\, .\label{fullv}
\end{equation}
In the NLL approximation it is necessary
to compute $\alpha^{(2)}(t)$ and $C^{gg(1)}$; however, to one loop
only $C^{gg(1)}$ appears.  In addition, only the dispersive parts of 
the one-loop amplitude contribute at NLL.  Expanding  eq.~(\ref{elasb}) 
to $O(g^4)$ and using eq.~(\ref{elas}), we obtain
\begin{eqnarray}
 {\rm Disp}\,M^{aa'bb'}_{\nu_a\nu_{a'}\nu_{b'}\nu_b}&=& M_4^{\rm tree}\Biggl\{
1\ +\ g^2 \Biggl[\alpha^{(1)}(t) \ln{s\over -t}\ +\ 
{{\rm Disp}\,C^{gg(1)}_{-\nu_a\nu_{a'}}(-p_a,p_{a'})\over 
C^{gg(0)}_{-\nu_a\nu_{a'}}(-p_a,p_{a'})} 
\nonumber\\ 
&&\qquad\qquad\qquad\quad +\ 
{{\rm Disp}\,C^{gg(1)}_{-\nu_b\nu_{b'}}(-p_b,p_{b'})\over
C^{gg(0)}_{-\nu_b\nu_{b'}}(-p_b,p_{b'})}\Biggr]\Biggr\}
\, .\label{expa}
\end{eqnarray}
Thus, the NLL corrections to ${\rm Disp}\, C^{gg(1)}$ can be extracted
directly from the 1-loop $g\,g\to g\,g$ amplitude.
We shall compute these virtual corrections to the vertices (\ref{centrc}) 
in sect.~\ref{sec:b}.

Finally, the NLO impact factors are obtained by combining the
square of the vertices $g^*\, g \rightarrow g\, g$ (\ref{treposb})
and $g^*\, g \rightarrow \bar q q$ (\ref{cqqa}), integrated over the
phase space of the intermediate gluon, with 
the gluon-loop and quark-loop contributions to ${\rm Disp}\,
C^{gg(1)}$, respectively.

\section{The 1-loop corrections to the $C$ vertices}
\label{sec:b}

\subsection{The 1-loop four-gluon amplitude}
\label{sec:ba}

The color decomposition of a tree-level multigluon
amplitude in a helicity basis is \cite{mp}
\begin{equation}
M_n^{tree} = 2^{n/2}\, g^{n-2}\, \sum_{S_n/Z_n} {\rm tr}(\lambda^{d_{\sigma(1)}} 
\cdots
\lambda^{d_{\sigma(n)}}) \, m_n(p_{\sigma(1)},\nu_{\sigma(1)};...;
p_{\sigma(n)},\nu_{\sigma(n)})\, ,\label{one}
\end{equation}
where $d_1,..., d_n$, and $\nu_1,..., \nu_n$ are
respectively the colors and the
polarizations of the gluons, the $\lambda$'s are the color 
matrices\footnote{Note that 
eq.(\ref{one}) differs by the $2^{n/2}$ factor from the expression
given in ref.\cite{mp}, because we use the standard normalization of
the $\lambda$ matrices, ${\rm tr}(\lambda^a\lambda^b) =
\delta^{ab}/2$.} in the
fundamental representation of SU($N_c$) and the sum is over the noncyclic
permutations $S_n/Z_n$ of the set $[1,...,n]$. We take
all the momenta as outgoing, and consider the {\sl maximally helicity-violating}
configurations $(-,-,+,...,+)$ for which the gauge-invariant subamplitudes,
$m_n(p_1,\nu_1; ...; p_n,\nu_n)$, assume the form \cite{mp},
\begin{equation}
m_n(-,-,+,...,+) = {\langle p_i p_j\rangle^4\over
\langle p_1 p_2\rangle \cdots\langle p_{n-1} p_n\rangle 
\langle p_n p_1\rangle}\, ,\label{two}
\end{equation}
where $i$ and $j$ are the gluons of negative helicity. The configurations
$(+,+,-,...,-)$ are then obtained by replacing the $\langle p k\rangle$
products with $\left[k p\right]$ products.  We give the formulae for these
spinor products in appendix A.  Using the high-energy limit
of the spinor products (\ref{hpro}), the tree-level amplitude for 
$g\,g\to g\,g$ scattering may be cast in the form (\ref{elas}).

The color decomposition of one-loop multigluon amplitudes is also
known \cite{bk1}. For four gluons it is,
\begin{eqnarray}
M_4^{1-loop} &=& 4 g^4 \left[\sum_{S_4/Z_4} \,{\rm tr}
(\lambda^{d_{\sigma(1)}} 
\lambda^{d_{\sigma(2)}} \lambda^{d_{\sigma(3)}} 
\lambda^{d_{\sigma(4)}}) \, m_{4:1}(\sigma(1), \sigma(2),
\sigma(3), \sigma(4)) \right. \nonumber\\
&&+ \left. \sum_{S_4/Z_2^3} {\rm tr}(\lambda^{d_{\sigma(1)}} 
\lambda^{d_{\sigma(2)}}) {\rm tr}(\lambda^{d_{\sigma(3)}} 
\lambda^{d_{\sigma(4)}})\, m_{4:3}(\sigma(1), \sigma(2),
\sigma(3), \sigma(4)) \right]\, ,\label{loop}
\end{eqnarray}
where  $\sigma(i)$ is a 
shorthand for $p_{\sigma(i)},\nu_{\sigma(i)}$ in the subamplitudes.
In the second line the sum is over the permutations of the four color
indices, up to permutations within each trace and to permutations
which interchange the two traces. There are two independent 
helicity-conserving configurations $(-,-,+,+)$ and $(-,+,-,+)$, 
for which the subamplitudes of the type $m_{4:1}$ are \cite{bk2}, 
\cite{kst},
\begin{eqnarray}
\lefteqn{ m_{4:1}(-,-,+,+) = m_4(-,-,+,+)\, c_{\Gamma}}\nonumber\\
& &\quad \times\left\{ \left(-{\mu^2\over s_{14}}\right)^{\epsilon} 
\left[N_c \left(
-{4\over\epsilon^2} - {11\over 3\epsilon} + {2\over\epsilon}
\ln{s_{12}\over s_{14}} - {64\over 9} - {\delta_R\over 3} + \pi^2 \right)
\right.\right.\nonumber\\
&&\left.\left.\quad
+ N_f
\left( {2\over 3\epsilon} + {10\over 9}\right)\right] - {\beta_0
\over \epsilon}\right\} \label{four}\\ & &\nonumber\\
\lefteqn{ m_{4:1}(-,+,-,+) = m_4(-,+,-,+)\, c_{\Gamma}}\nonumber\\
& &\quad \times\left\{ \left(-{\mu^2\over s_{14}}\right)^{\epsilon} 
\left[ N_c\left(
-{4\over\epsilon^2} - {11\over 3\epsilon} + {2\over\epsilon}
\ln{s_{12}\over s_{14}} - {64\over 9} - {\delta_R\over 3} + \pi^2 \right)
+ N_f\left( {2\over 3\epsilon} + 
{10\over 9}\right)\right.\right. \nonumber\\
& &\quad -(N_c-N_f)\, {s_{12}s_{14}\over s_{13}^2} 
\left[1-\left({s_{12}\over s_{13}} - {s_{14}\over s_{13}}\right) 
\ln{s_{14}\over s_{12}} - \left({s_{12}s_{14}\over s_{13}^2} - 2\right) 
\left(\ln^2{s_{14}\over s_{12}} + \pi^2\right)\right]\nonumber\\
& & \left.\left.\quad + \beta_0 
{s_{14}\over s_{13}} \ln{s_{14}\over s_{12}}
- {3\over 2} N_f 
{s_{12}s_{14}\over s_{13}^2} \left(\ln^2{s_{14}\over
s_{12}} + \pi^2\right) \right]
- {\beta_0\over \epsilon}\right\}\, ,\label{five}
\end{eqnarray}
with $N_f$ the number of quark flavors, $c_{\Gamma}$ given in 
eq.~(\ref{cgam}), $\beta_0 = (11N_c-2N_f)/3$,
\begin{equation}
\delta_R = \left\{ \begin{array}{ll} 1 & \mbox{HV or CDR scheme},\\
0 & \mbox{dimensional reduction scheme}, \end{array} \right. \label{cp}
\end{equation}
the tree amplitude $m_4$ given in eq.~(\ref{two}), and the
$\overline{\rm MS}$ ultraviolet counterterm in the last term of eq.~(\ref{four})
and (\ref{five}).

There are three subamplitudes of the type $m_{4:3}$ to
be determined in the second line of eq.~(\ref{loop}):
$m_{4:3}(1,2,3,4)$, $m_{4:3}(1,3,2,4)$ and $m_{4:3}(1,4,2,3)$.
However, any subamplitude of the type $m_{4:3}$ may be
obtained from the subamplitudes of type $m_{4:1}$~\cite{bk1}.
They satisfy
\begin{equation}
m_{4:3}(1,2,3,4) = {1\over N_c}\, 
\sum_{S_4/Z_4} m_{4:1}(1,2,3,4)\ ,\label{nonp}
\end{equation}
where only the $N_f$-independent, unrenormalized contributions to
$m_{4:1}$ are included in this formula~\cite{kst,bdk5pt}.
Then we have
\begin{equation}
m_{4:3}(1,2,3,4) = m_{4:3}(1,3,2,4) = m_{4:3}(1,4,2,3)
\ ,\label{equal}
\end{equation}
and it suffices to determine $m_{4:3}(1,2,3,4)$. 

To obtain the next-to-leading log corrections to the helicity-conserving 
$g^*\, g\to g$ vertex, we need the amplitude 
$M_4^{1-loop}(B-,A-,A'+,B'+)$ in the high-energy limit.
We must consider each of the color orderings in eq.~(\ref{loop})
and expand the expressions (\ref{four}) and (\ref{five}) in 
powers of $t/s$, retaining only the leading power, which yields
the leading and next-to-leading terms in $\ln(s/t)$.
In fact, at next-to-leading log, we only need to keep the dispersive 
parts of the subamplitudes $m_{4:1}$ and $m_{4:3}$.  It
is not difficult to show that if a given color ordering of $m_{4}$ 
is suppressed by a power of $t/s$ at tree-level, then the
corresponding color ordering of $m_{4:1}$ will also
be suppressed at one-loop.  Then the leading color orderings of type 
$m_{4:1}(-,-,+,+)$ (as in eq.~(\ref{four})) 
occur with $s_{12}=s , s_{14}=t$. 
Thus, we have
\begin{eqnarray}
\lefteqn{ {\rm Disp}\, m_{4:1}(-,-,+,+) = m_4(-,-,+,+)\, 
c_{\Gamma}}\nonumber\\
& &\qquad\qquad \times\left\{ \left({\mu^2\over -t}\right)^{\epsilon} \left[ 
N_c\left(
-{4\over\epsilon^2} - {11\over 3\epsilon} + {2\over\epsilon}
\ln{s\over -t} - {64\over 9} - {\delta_R\over 3} + \pi^2 \right)
\right.\right.\nonumber\\
&&\left.\left.\qquad\qquad+ N_f
\left( {2\over 3\epsilon} + {10\over 9}\right)\right] - {\beta_0
\over \epsilon}\right\}\, .\label{six}
\end{eqnarray}
The leading color orderings of type
$m_{4:1}(-,+,-,+)$ (as in eq.~(\ref{five})) occur with $s_{14}=t, 
s_{12}=u$ or $s_{14}=u , s_{12}=t$.  However, eq.~(\ref{five}) is
symmetric in $s_{14}$ and $s_{12}$ up to ${\cal O}(\epsilon)$, so both
orderings give the same result\footnote{Using the reflection and
cyclic symmetries of the subamplitudes~\cite{bk1}, 
we see that $m_{4:1}(A-,A'+,B-,B'+)=m_{4:1}(A-,B'+,B-,A'+)$,
i.e. if eq.~(\ref{five}) were calculated to all orders in
$\epsilon$ it would be exactly symmetric in $s_{14}$ and $s_{12}$.}. 
Using $u=-s-t$, we see that eq.~(\ref{six})
holds for the dispersive parts of $m_{4:1}(-,+,-,+)$ as well.
Finally, the subamplitude $m_{4:3}$ can be obtained
using eq.~(\ref{six}) and (\ref{nonp}).
We find that Disp~$m_{4:3}$ vanishes to power accuracy in $s/t$:
\begin{equation}
{\rm Disp}\, m_{4:3}(B-,A-,A'+,B'+) = 0 + O(t/s)\, .\label{simp}
\end{equation}
The other color orderings of $m_{4:3}$ also vanish due to (\ref{equal}).
Thus, we conclude that the dispersive part of the one-loop amplitude
is simply proportional to the tree amplitude to leading power in $t/s$:
\begin{eqnarray}
\lefteqn{ {\rm Disp}\, M_4^{1-loop}(B-,A-,A'+,B'+) = 
M_4^{tree}(B-,A-,A'+,B'+)\, g^2\, c_{\Gamma}}\nonumber\\
& & \qquad
\qquad\times\left\{ \left({\mu^2\over -t}\right)^{\epsilon} \left[ N_c\left(
-{4\over\epsilon^2} - {11\over 3\epsilon} + {2\over\epsilon}
\ln{s\over -t} - {64\over 9} - {\delta_R\over 3} + \pi^2 \right)
\right.\right.\nonumber\\
&&\left.\left.\qquad\qquad
+ N_f\left( {2\over 3\epsilon} + {10\over 9}\right)\right] - {\beta_0
\over\epsilon}\right\}\, .\label{seven}
\end{eqnarray}

To LL accuracy, using eq.~(\ref{alph}), 
we find that eq.~(\ref{seven}) reduces to
\begin{equation}
{\rm Disp}\, M_4^{1-loop}(B-,A-,A'+,B'+) = g^2 \alpha^{(1)}(t)\, 
\ln{s\over -t}\, M_4^{\rm tree}\, ,\label{ll}
\end{equation}
in agreement with eq.~(\ref{expa}). To NLL accuracy, we may extract
from eq.~(\ref{expa}) and (\ref{seven}) the 1-loop corrections
to the helicity-conserving vertex\footnote{Of course,
there is no gauge-invariant way of distinguishing the contribution
to $C^{gg(1)}_{-\nu_a\nu_{a'}}(-p_a,p_{a'})$ from the one to
$C^{gg(1)}_{-\nu_b\nu_{b'}}(-p_b,p_{b'})$, so we conventionally assume
that they are equal.} $g^*\, g\to g$,
\begin{eqnarray}
\lefteqn{{{\rm Disp}\, C^{gg(1)}_{-+}(-p_a,p_{a'})\over 
C^{gg(0)}_{-+}(-p_a,p_{a'})} =
{{\rm Disp}\, C^{gg(1)}_{-+}(-p_b,p_{b'})\over
C^{gg(0)}_{-+}(-p_b,p_{b'})} =} \nonumber\\ 
&& \quad c_{\Gamma} \left\{ \left({\mu^2\over -t}\right)^{\epsilon} 
\left[ N_c\left(-{2\over\epsilon^2} - {11\over 6\epsilon} 
- {32\over 9} - {\delta_R\over 6} + {\pi^2\over 2} \right)
+ N_f\left( {1\over 3\epsilon} + {5\over 9}\right)\right] - {\beta_0
\over 2\epsilon}\right\}\, .\label{vert}
\end{eqnarray}

We can compare this to the unrenormalized one-loop corrections to the
helicity-conserving vertex calculated in ref.~\cite{fl} 
and \cite{ff}\footnote{The purely gluonic part of the 
unrenormalized 1-loop corrections in ref.~\cite{fl} is
marred by misprints and mistakes. The correct expression is
given in ref.~\cite{lipa}.}. Rewriting it as
\begin{eqnarray}
 \Gamma_{-+}^{(1)}(t) &=& c_{\Gamma} \left({\mu^2\over -t}
\right)^{\epsilon} {1\over\epsilon (1-2\epsilon)}
\left\{ N_f {1-\epsilon\over 3-2\epsilon}
\right. \nonumber\\ & & \left. +
N_c\left[ (1-2\epsilon) [\psi(1+\epsilon) -2\psi(-\epsilon)
+\psi(1)] - {1\over 4 (3-2\epsilon)} -
{7\over 4}\right] \right\}\, ,\label{fffl}
\end{eqnarray}
and expanding to $O(\epsilon^0)$, we see that it agrees with 
eq.~(\ref{vert}), with $\delta_R = 1$.

At one-loop there are also contributions to the helicity-violating
part of the vertex $C^{gg(1)}$.  To calculate this 
we need the subamplitudes \cite{bk2}, \cite{kst},
\begin{eqnarray}
m_{4:1}(-,+,+,+) &=& {1\over 48\pi^2}\, (N_c-N_f)\,
{[24]^2\over [12] \langle 23\rangle \langle 34\rangle
[41]}\, (s_{12}+s_{14}) \label{viola}\\ 
m_{4:3}(-,+,+,+) &=& {1\over 8\pi^2}\,
{\langle 12\rangle [24] \langle 41\rangle \over 
\langle 23\rangle \langle 34\rangle \langle 24\rangle}\,
.\label{violb}
\end{eqnarray}
Using eq.~(\ref{viola}) and the spinor products (\ref{hpro}),
we find in the high-energy limit,
\begin{eqnarray}
\lefteqn{ m_{4:1}\left(\sigma(B-),\sigma(A+),\sigma(A'+),\sigma(B'+)\right)
}\nonumber\\ & & \qquad= - {1\over 48\pi^2}\, (N_c-N_f)\, 
{p_{b'\perp}^*\over p_{b'\perp}}\, m_4\left(\sigma(B-),\sigma(A-),
\sigma(A'+),\sigma(B'+)\right)\, ,\label{helv}
\end{eqnarray}
where $\left(\sigma(B),\sigma(A),\sigma(A'),\sigma(B')\right)$
spans the four permutations:
$(B,A,A',B')$, $(B,B',A,A')$, $(B,A',A,B')$ and $(B,B',A',A)$,
which yield the leading color orderings at tree level \cite{ptlip}.
The other color orderings are subleading and, using eq.~(\ref{violb}),
we find that $m_{4:3}(-,+,+,+)$
is also subleading in the high-energy limit. Therefore we have,
\begin{equation}
M_4^{1-loop}(B-,A+,A'+,B'+) = - {g^2\over 48\pi^2}\, (N_c-N_f)\,
{p_{b'\perp}^*\over p_{b'\perp}}\, M_4^{tree}(B-,A-,A'+,B'+)\,
.\label{helviol}
\end{equation}
From eq.~(\ref{elas}), and (\ref{elasb}) expanded to $O(g^4)$,
with $C^{gg(0)}_{++} = 0$, we may write,
\begin{equation}
M_4^{1-loop}(B-,A+,A'+,B'+) = g^2 {C^{gg(1)}_{++}(-p_a,p_{a'})\over 
C^{gg(0)}_{-+}(-p_a,p_{a'})}\, M_4^{tree}(B-,A-,A'+,B'+)\,
,\label{ten}
\end{equation}
and comparing eq.~(\ref{ten}) to (\ref{helviol}), we obtain
\begin{equation}
{C^{gg(1)}_{++}(-p_a,p_{a'})\over C^{gg(0)}_{-+}(-p_a,p_{a'})} =
- {1\over 48\pi^2}\, (N_c-N_f)\, {p_{b'\perp}^*\over p_{b'\perp}}\,
.\label{elev}
\end{equation}

Using the amplitude $M_4^{1-loop}(B-,A-,A'-,B'+)$, it is possible
to check that the 1-loop corrections maintain the property
of complex conjugation under helicity reversal, i.e.
that $[C^{gg(1)}_{++}]^* = C^{gg(1)}_{--}$.
The helicity-violating part of the vertex has been also
computed in ref.\cite{fl}, \cite{ff}
\begin{equation}
\Gamma_{++}^{(1)}(t) = c_{\Gamma} \left({\mu^2\over -t}\right)^{\epsilon} 
{1\over (3-2\epsilon)(1-2\epsilon)} \left(N_c - {N_f \over 1-\epsilon}
\right)\, ,\label{twel}
\end{equation}
which, when expanded to $O(\epsilon^0)$, coincides with eq.~(\ref{elev})
up to an irrelevant overall phase.

\subsection{The 1-loop four-quark amplitude}
\label{sec:c}

The tree-level amplitude for $q\,q\to q\,q$ scattering 
in the high-energy limit is
\begin{equation}
M^{\bar aa'\bar bb'\,{\rm tree}}_{\nu_a\nu_b} = 2 s \left[g\, 
\lambda^c_{a' \bar a}\,
C_{-\nu_a \nu_a}^{\bar q q(0)}(-p_a,p_{a'}) \right] {1\over t} 
\left[g\, \lambda^c_{b' \bar b}\,
C_{-\nu_b \nu_b}^{\bar q q(0)}(-p_b,p_{b'}) \right]
,\label{elasq}
\end{equation}
where we have labelled the incoming quarks
as outgoing antiquarks by convention. 
The $C$-vertices $g^*\, q \rightarrow q$ are
\begin{equation}
C_{-+}^{\bar q q(0)}(-p_a,p_{a'}) = -1\, ;\qquad 
C_{-+}^{\bar q q(0)}(-p_b,p_{b'}) = - \left({p_{b'\perp}^* 
\over p_{b'\perp}}\right)^{1/2}\, ,\label{cbqqm}
\end{equation}
and we note that helicity is exactly conserved on a 
massless fermion line. The tree-level high-energy limit for
quark-quark scattering (\ref{elasq}) is similar to
that for gluon-gluon scattering (\ref{elas}), since both of 
them are dominated by gluon exchange in the $t$-channel. This 
suggests that the virtual radiative corrections to eq.~(\ref{elasq}) in
the LL approximation can be obtained by Reggeizing the $t$-channel gluon 
as in eq.~(\ref{sud}), and that the general form of the quark-quark 
scattering amplitude in the high-energy limit is essentially the same
as that for gluon-gluon scattering (\ref{elasb}).  Accordingly, its 
expansion to $O(g^4)$ yields
\begin{eqnarray}
{\rm Disp}\,M^{\bar aa'\bar bb'}_{\nu_a\nu_b} &=& M_4^{\rm 
tree}\Biggl\{1\ +\ g^2 \Biggl[\alpha^{(1)}(t)\ln{s\over -t}\ +\ 
{{\rm Disp}\,C_{-\nu_a \nu_a}^{\bar q q(1)}(-p_a,p_{a'})\over
C_{-\nu_a \nu_a}^{\bar q q(0)}(-p_a,p_{a'})} 
\nonumber\\ 
& &\qquad\qquad\qquad\quad+\ 
{{\rm Disp}\,C_{-\nu_b \nu_b}^{\bar q q(1)}(-p_b,p_{b'})\over
C_{-\nu_b \nu_b}^{\bar q q(0)}(-p_b,p_{b'})}
\Biggr]\Biggr\}
\, ,\label{expq}
\end{eqnarray}
with $M_4^{\rm tree}$ in eq.~(\ref{elasq}).

The color decomposition of the quark-quark scattering amplitude is
respectively at tree level \cite{mp}
\begin{equation}
M_4^{tree}(\bar q,\bar Q; Q, q) = g^2 \left(\delta_{i_1i_3}
\delta_{i_2i_4} - {1\over N_c} \delta_{i_1i_4} \delta_{i_2i_3}
\right) a_4(1, 2; 3, 4)\, ,\label{qtree}
\end{equation}
and at one loop \cite{kst}
\begin{eqnarray}
M_4^{1-loop}(\bar q,\bar Q; Q, q) &=& g^4 \left[\left(\delta_{i_1i_3}
\delta_{i_2i_4} - {1\over N_c} \delta_{i_1i_4} \delta_{i_2i_3}
\right) a_{4:1}(1, 2; 3, 4) \right. \nonumber\\ &+& 
\left. \delta_{i_1i_3} \delta_{i_2i_4} a_{4:2}(1, 2; 3, 4) \right]\, 
.\label{qloop}
\end{eqnarray}
The tree-level subamplitude is
\begin{equation}
a_4(-, -; +, +) = {\langle 12\rangle [34]\over s_{14}}\,
.\label{subt}
\end{equation}
Using the spinor products (\ref{hpro}) in the high-energy limit
in this equation and substituting into eq.~(\ref{qtree}), we obtain
eq.~(\ref{elasq}).

The one-loop subamplitudes are \cite{kst}
\begin{eqnarray}
\lefteqn{a_{4:1}(-, -; +, +) = a_4(-, -; +, +) c_{\Gamma}}
\nonumber\\ & &\quad \times\left\{ \left(-{\mu^2\over s_{14}}
\right)^{\epsilon} \left[ N_c\left(-{2\over\epsilon^2} + 
{2\over 3\epsilon} + {2\over\epsilon} \ln{s_{12}\over s_{14}} 
+ {19\over 9} - {2\delta_R\over 3} + \pi^2 \right)
- N_f\left( {2\over 3\epsilon} + {10\over 9}\right) 
\right.\right. \nonumber\\ & &\quad + {1\over N_c} 
\left({2\over\epsilon^2} + {3\over\epsilon} +
{2\over\epsilon} \ln{s_{12}\over s_{13}} + 7 + \delta_R
- {1\over 2} {s_{14}\over s_{12}} \left(1 - 
{s_{13}\over s_{12}}\right) \left( \ln^2{s_{14}\over 
s_{13}} + \pi^2\right) \right. \nonumber\\ & &\quad 
\left.\left.\left. - {s_{14}\over s_{12}}
\ln{s_{14}\over s_{13}}\right) \right] - {\beta_0
\over\epsilon}\right\} \label{pippo}\\ & &\nonumber\\
\lefteqn{a_{4:2}(-, -; +, +) = a_4(-, -; +, +) c_{\Gamma}} 
\nonumber\\ & &\quad \times \left(-{\mu^2\over s_{14}}
\right)^{\epsilon} {N_c^2-1\over N_c} \left[ -
{2\over\epsilon} \ln{s_{12}\over s_{13}}
+ {1\over 2} {s_{14}\over s_{12}} \left(1 - 
{s_{13}\over s_{12}}\right) \left( \ln^2{s_{14}\over 
s_{13}} + \pi^2\right)\right. \nonumber\\
&&\quad\left.+ {s_{14}\over s_{12}}
\ln{s_{14}\over s_{13}}\right]\, .\label{pippa} 
\end{eqnarray}
As in sect.~\ref{sec:ba}, we only consider the dispersive part of
these equations.  Expanding in powers of $t/s$ and 
retaining only the leading power yields
the LL and NLL terms in $\ln(s/t)$:
\begin{eqnarray}
\lefteqn{{\rm Disp}\, a_{4:1}(B-,A-;A'+,B'+) = 
a_4(B-,A-;A'+,B'+) c_{\Gamma}}
\nonumber\\ & &\qquad \times\left\{ \left(-{\mu^2\over t}
\right)^{\epsilon} \left[ N_c\left(-{2\over\epsilon^2} + 
{2\over 3\epsilon} + {2\over\epsilon} \ln{s\over -t} 
+ {19\over 9} - {2\delta_R\over 3} + \pi^2 \right)
\right.\right. \nonumber\\ & &\qquad \left.\left.
- N_f\left( {2\over 3\epsilon} + {10\over 9}\right)
 + {1\over N_c} 
\left({2\over\epsilon^2} + {3\over\epsilon} + 7 + \delta_R
\right) \right] - {\beta_0\over\epsilon}\right\}\,
.\label{hepip}
\end{eqnarray}
The dispersive part of $a_{4:2}$ is subleading to power accuracy, 
and we neglect it. Substituting eq.~(\ref{hepip}) into
eq.~(\ref{qloop}) and using eq.~(\ref{qtree}), we obtain
for the configuration $(B-,A-;A'+,B'+)$:
\begin{eqnarray}
\lefteqn{{\rm Disp}\, M_4^{1-loop}(\bar q,\bar Q, Q, q) =
M_4^{tree}(\bar q,\bar Q, Q, q) g^2 c_{\Gamma}}
\nonumber\\ & &\qquad \times\left\{ \left(-{\mu^2\over t}
\right)^{\epsilon} \left[ N_c\left(-{2\over\epsilon^2} + 
{2\over 3\epsilon} + {2\over\epsilon} \ln{s\over -t} 
+ {19\over 9} - {2\delta_R\over 3} + \pi^2 \right)
\right.\right. \nonumber\\ & &\qquad \left.\left.
- N_f\left( {2\over 3\epsilon} + {10\over 9}\right)
 + {1\over N_c} 
\left({2\over\epsilon^2} + {3\over\epsilon} + 7 + \delta_R
\right) \right] - {\beta_0\over\epsilon}\right\}\, 
.\label{dispq}
\end{eqnarray}
To LL accuracy,
eq.~(\ref{dispq}) reduces to 
\begin{equation}
{\rm Disp}\, M_4^{1-loop}(B-,A-;A'+,B'+) = g^2 \alpha^{(1)}(t)\, 
\ln{s\over -t}\, M_4^{\rm tree}\, ,\label{llqq}
\end{equation}
which is exactly the same form as eq.~(\ref{ll}),
due to the universality of the LL contribution.
Comparing the expansion (\ref{expq}) to eq.~(\ref{dispq}), we 
obtain the 1-loop
correction to the $C$-vertex $g^*\, q \rightarrow q$,
\begin{eqnarray}
\lefteqn{{{\rm Disp}\, C_{-+}^{\bar q q(1)}(-p_a,p_{a'})\over
C_{-+}^{\bar q q(0)}(-p_a,p_{a'})} =
{{\rm Disp}\, C_{-+}^{\bar q q(1)}(-p_b,p_{b'})\over
C_{-+}^{\bar q q(0)}(-p_b,p_{b'})} = }\nonumber\\
& &\qquad c_{\Gamma} \left\{ \left(-{\mu^2\over t}\right)^{\epsilon} 
\left[ N_c\left(-{1\over\epsilon^2} + {1\over 3\epsilon}
+ {19\over 18} - {\delta_R\over 3} + {\pi^2\over 2} \right)
\right.\right. \nonumber\\ & &\qquad \left.\left.
- N_f\left( {1\over 3\epsilon} + {5\over 9}\right)
 + {1\over N_c} 
\left({1\over\epsilon^2} + {3\over 2\epsilon} + {7\over 2} 
+ {\delta_R\over 2}\right) \right] - {\beta_0\over 2\epsilon}
\right\}\, .\label{lqver}
\end{eqnarray}
The unrenormalized 1-loop corrections to the $C$-vertex 
$g^*\, q \rightarrow q$ have also been computed in ref.~\cite{ffq},
\begin{eqnarray}
\Gamma_{-+\, q}^{(1)}(t) &=& c_{\Gamma} \left({\mu^2\over -t}
\right)^{\epsilon} {1\over\epsilon (1-2\epsilon)}
\left\{ - N_f {1-\epsilon\over 3-2\epsilon} + {1\over N_c}
\left({1\over\epsilon} - {1-2\epsilon\over 2}\right)
\right. \nonumber\\ & & \left. +
N_c\left[ (1-2\epsilon) [\psi(1+\epsilon) -2\psi(-\epsilon)
+\psi(1)] + {1\over 4 (3-2\epsilon)} + {1\over\epsilon} -
{7\over 4}\right] \right\}\, .\label{ffqvert}
\end{eqnarray}
Expanding this to $O(\epsilon^0)$, we see that 
it agrees with eq.~(\ref{lqver}), with $\delta_R = 1$.

\section{Conclusions}
\label{sec:conclusions}

Our main results are the 1-loop corrections to 
the $g^*\, g \rightarrow g$ vertex, eqs.~(\ref{vert}) and (\ref{elev}), and 
to the $g^*\, q \rightarrow q$ vertex, eq.~(\ref{lqver}).  
The dispersive parts, which are all that is needed at NLL, agree with 
previous calculations performed in the CDR scheme \cite{lipa}, 
which have been obtained in a different 
manner.  We also give the absorptive parts in appendix C.  The
virtual corrections, when combined with the $g^{*}\,g\rightarrow g\,g$
(eq.~(\ref{nllfg})), $g^{*}\,g\rightarrow q\,\bar q$
(eq.~(\ref{forwqq})), and $g^{*}\,q\rightarrow g\,q$
\cite{thuile} real corrections, give the NLO gluon and quark impact factors,
which are necessary to use the BFKL resummation at NLL for jet
production at both lepton-hadron and hadron-hadron colliders.

The NLL corrections to the BFKL approximation are greatly desired, 
because they incorporate several physical effects which are lacking in 
the LL resummation.  For instance, the real corrections to the impact 
factor contain a collinear singularity, as explicitly seen in 
eq.~(\ref{treposb}).  Thus, it is possible only beyond LL to consider 
effects such as jet definition and cone-size dependence of the cross 
section.  In addition, the NLL corrections also improve the 
errors due to energy and longitudinal momentum conservation.  
Although these effects can be included in the parton density 
functions through a BFKL Monte Carlo simulation \cite{carl,os}, there are 
still large uncertainties due to the LL approximation used for the 
matrix elements.  The NLL corrections should reduce these 
uncertainties substantially.

Finally, the incorporation of renormalization- and 
factorization-scale dependence only becomes manifest at NLL.  Note 
that the virtual correction to the $g^{*}\,g\rightarrow g$ vertex 
(\ref{vert}) can be written
\begin{equation}
{{\rm Disp}\, C^{gg(1)}_{-+}\over 
C^{gg(0)}_{-+}}\  =\ 
c_{\Gamma} \left({\mu^2\over -t}\right)^{\epsilon} 
\left[-{1\over2}\beta_{0}\ln{-t\over\mu^{2}}\ +\ 
\mbox{non-log terms}\right]\, .\label{scale}
\end{equation}
Thus, the scale dependence of the amplitude can be obtained by simply 
replacing $g(\mu^{2})\rightarrow 
g(-t)=g(\mu^{2})\left[1-(\beta_{0}\,\alpha_{s}/8\pi)\ln(-t/\mu^2)
+\dots\right]$.
Of course, this is to be expected, since there is only one scale, 
$-t\simeq q_{\perp}^{2}$, which enters the impact factor.  However, in 
the BFKL kernel, there are several scales which enter the vertex 
$g^{*}(q_{\perp})\,g^{*}(q_{\perp}+k_{\perp})\rightarrow 
g(k_{\perp})$.  At LL these scales are all assumed comparable, and 
cannot be distinguished as far as the choice of renormalization scale.
By studying the NLL corrections to this vertex \cite{fl,noi}, which
can be obtained 
from the 1-loop five-gluon amplitudes \cite{bdk5pt}, one can 
investigate the proper scale-dependence of the BFKL kernel.

\appendix
\section{Spinor Algebra in the Multi-Regge kinematics}
\label{sec:appa}

We consider the scattering of two gluons of momenta $p_a$ and $p_b$
into $n+2$ gluons of momenta $p_{i}$, where $i=a',b',1\dots n$.
 Using light-cone coordinates $p^{\pm}= p_0\pm p_z$, and
complex transverse coordinates $p_{\perp} = p_x + i p_y$, with scalar
product $2 p\cdot q = p^+q^- + p^-q^+ - p_{\perp} q^*_{\perp} - p^*_{\perp} 
q_{\perp}$, the gluon 4-momenta are,
\begin{eqnarray}
p_a &=& \left(p_a^+, 0; 0, 0\right)\, ,\nonumber \\
p_b &=& \left(0, p_b^-; 0, 0\right)\, ,\label{in}\\
p_i &=& \left(|p_{i\perp}| e^{y_i}, |p_{i\perp}| e^{-y_i}; 
|p_{i\perp}|\cos{\phi_i}, |p_{i\perp}|\sin{\phi_i}\right)\, ,\nonumber
\end{eqnarray}
where to the left of the semicolon we have the + and -
components, and to the right the transverse components.
$y$ is the gluon rapidity and $\phi$ is the azimuthal angle between the 
vector $p_{\perp}$ and an arbitrary vector in the transverse plane.
Momentum conservation gives
\begin{eqnarray}
0 &=& \sum p_{i\perp}\, ,\nonumber \\
p_a^+ &=& \sum p_{i}^+ \, ,\label{kin}\\ 
p_b^- &=& \sum p_{i}^- \, .\nonumber
\end{eqnarray}

For each massless momentum $p$ there is a positive and negative helicity 
spinor, $|p+\rangle$ and $|p-\rangle$, so we can consider two types of 
spinor products 
\begin{eqnarray}
        \langle pq\rangle & = & \langle p-|q+\rangle\nonumber\\
        \left[ pq \right] & = & \langle p+|q-\rangle\ .
        \label{spinors}
\end{eqnarray}
Phases are chosen so that $\langle pq\rangle=-\langle qp\rangle$
and $[pq]=-[qp]$.
For the momentum under consideration the spinor products are
\begin{eqnarray}
\langle p_{i} p_{j}\rangle &=& p_{i\perp}\sqrt{p_{j}^+
\over p_{i}^+} - p_{j\perp} \sqrt{p_{i}^+\over p_{j}^+}\, 
,\nonumber\\ \langle p_a p_i\rangle &=& -\sqrt{p_a^+
\over p_i^+}\, p_{i\perp}\, ,\label{spro}\\ \langle p_i p_b\rangle &=&
-\sqrt{p_i^+ p_b^-}\, ,\nonumber\\ \langle p_a p_b\rangle 
&=& - \sqrt{p_a^+ p_b^-} = -\sqrt{s_{ab}}\, ,\nonumber
\end{eqnarray}
where we have used the mass-shell condition 
$|p_{i\perp}|^2 = p_i^+ p_i^-$.  The other type of spinor product 
can be obtained from
\begin{equation}
        [pq]\ =\ \pm\langle qp\rangle^{*}\ ,
\end{equation}
where the $+$ is used if $p$ and $q$ are both ingoing or both 
outgoing, and the $-$ is used if one is ingoing and the other 
outgoing.

In the multi-Regge kinematics, the gluons are strongly ordered in 
rapidity and have comparable transverse momentum:
\begin{equation}
y_{a'} \gg y_1\gg\dots y_{n} \gg y_{b'};\qquad |p_{i\perp}|\simeq|p_{\perp}|\, 
.\nonumber
\end{equation}
Then the momentum conservation (\ref{kin}) in the $\pm$ directions reduces to
\begin{eqnarray}
p_a^+ &\simeq& p_{a'}^+\, ,\nonumber\\ 
p_b^- &\simeq& p_{b'}^-\, ,\label{hkin}
\end{eqnarray}
and the Mandelstam invariants become
\begin{eqnarray}
s_{ab} &=& 2 p_a\cdot p_b \simeq p_{a'}^+ p_{b'}^- \nonumber\\ 
s_{ai} &=& -2 p_a\cdot p_i \simeq - p_{a'}^+ p_i^- \label{mrinv}\\ 
s_{bi} &=& -2 p_b\cdot p_i \simeq - p_i^+ p_{b'}^- \nonumber\\ 
s_{ij} &=& 2 p_i\cdot p_j \simeq |p_{i\perp}| |p_{j\perp}| e^{y_i-y_j}
=p_{i}^{+}p_{j}^{-}\qquad (y_{i}\gg y_{j})\,
,\nonumber
\end{eqnarray}
where $i,j=a',b',1\dots n$.  
In this limit the
 spinor products (\ref{spro}) become
\begin{eqnarray}
\langle p_a p_b\rangle &\simeq& \langle p_{a'} p_b\rangle \simeq
-\sqrt{p_{a'}^+\over p_{b'}^+} |p_{b'\perp}| \nonumber\\
\langle p_a p_{b'}\rangle &\simeq& \langle p_{a'} p_{b'}\rangle =
-\sqrt{p_{a'}^+\over p_{b'}^+}\, p_{b'\perp} \nonumber\\
\langle p_a p_{a'}\rangle &\simeq& - p_{a'\perp} \nonumber\\
\langle p_{b'} p_b\rangle &\simeq& - |p_{b'\perp}| \label{hpro}\\
\langle p_a p_i\rangle &\simeq& \langle p_{a'} p_i\rangle =
-\sqrt{p_{a'}^+\over p_i^+}\, p_{i\perp} \nonumber\\
\langle p_i p_b\rangle &\simeq& -\sqrt{p_i^+\over p_{b'}^+} 
|p_{b'\perp}| \nonumber\\ 
\langle p_i p_{b'}\rangle &\simeq& -\sqrt{p_i^+\over p_{b'}^+} 
p_{b'\perp} \nonumber\\ 
\langle p_i p_{j}\rangle &\simeq& -\sqrt{p_i^+\over p_{j}^+}\,
p_{j\perp} \qquad (y_{i}\gg y_{j})\ .\nonumber
\end{eqnarray}

\section{Stringy Decomposition of the One-loop Amplitude}
\label{sec:appb}

It is instructive to compute eq.~(\ref{seven}) also from the 
string-inspired decomposition of a one-loop gluonic amplitude \cite{bdk},
\begin{equation}
m_{4:1} = N_c A_4^g + \left(4N_c-N_f\right) A_4^f + 
\left(N_c-N_f\right) A_4^s\, ,\label{stri}
\end{equation}
with $A_4^g$ the $N=4$ supersymmetric multiplet, $A_4^f$ the
$N=1$ chiral multiplet, and $A_4^s$ the contribution of a complex
scalar,
\begin{equation}
A_4^x = c_{\Gamma}\, M_4^{tree}\, (V^x+F^x)\, \qquad\qquad x=g,f,s\,
.\label{dec}
\end{equation}
The functions obtained from the $N=4$ multiplet, $V^g$ and $F^g$, 
are the same for any color ordering
\begin{equation}
F^g = 0\, ,\qquad V^g = -{2\over\epsilon^2} \left[\left({\mu^2\over -s_{12}}
\right)^{\epsilon} + \left({\mu^2\over -s_{23}}\right)^{\epsilon}\right] +
\ln^2\left({-s_{12}\over -s_{23}}\right) + \pi^2\, .\label{vertica} 
\end{equation}
The other functions depend on the color ordering. For orderings
of type $m_{4:1}(-,-,+,+)$ we have
\begin{eqnarray}
&F^f = 0\, ,& V^f = -{1\over\epsilon} \left({\mu^2\over -s_{23}}
\right)^{\epsilon}- 2\, ,\label{vertic}\\
&F^s = 0\, ,&V^s = -{V^f\over 3} + {2\over 9}\, ;\nonumber
\end{eqnarray}
and for color orderings of type $m_{4:1}(-,+,-,+)$ we have
\begin{eqnarray}
V^f &=& -{1\over 2\epsilon} \left[\left({\mu^2\over -s_{12}}
\right)^{\epsilon} + \left({\mu^2\over -s_{23}}\right)^{\epsilon}
\right] - 2\, ,\nonumber\\ 
V^s &=& -{V^f\over 3} + {2\over 9}\, ,\label{verticb}\\
F^f &=& -{s_{12} s_{23}\over 2s_{13}^2}
\left[\ln^2\left({-s_{12}\over -s_{23}}\right) + \pi^2\right]
+ {s_{12}-s_{23}\over 2s_{13}} \ln\left({-s_{12}\over -s_{23}}\right)
\, ,\nonumber\\
F^s &=& -{s_{12} s_{23}\over s_{13}^2} \left[
-{s_{12} s_{23}\over s_{13}^2}
\left(\ln^2\left({-s_{12}\over -s_{23}}\right) + \pi^2\right)
+ {s_{12}-s_{23}\over s_{13}} \left(1+ {s_{13}^2\over 6 s_{12} s_{23}}
\right) \ln\left({-s_{12}\over -s_{23}}\right) + 1\right]\, .\nonumber
\end{eqnarray}
Replacing the functions (\ref{dec}-\ref{verticb}) into 
eq.~(\ref{stri}) for the different color orderings and retaining only
the leading powers in $t/s$, we obtain the unrenormalized amplitude
(\ref{seven}) with $\delta_R=0$. We note that 
to LL accuracy only the functions (\ref{vertica}) obtained from the 
$N=4$ multiplet contribute, while to NLL accuracy also the functions
obtained from the $N=1$ chiral multiplet and the complex scalar 
contribute.

\section{Absorptive parts of the 1-loop amplitudes in the high-energy limit}
\label{sec:appc}

The absorptive parts of the 1-loop amplitudes in the high-energy limit
can be obtained using the same techniques as in section 3.  For the
gluon-gluon amplitude we find
\begin{eqnarray}
\lefteqn{{\rm Absorp}\, M_4^{1-loop}(B-,A-,A'+,B'+) \ =\  
g^4\,{s\over t} c_{\Gamma}
\left({\mu^2\over -t}\right)^{\epsilon} N_c\,{2\over\epsilon}
\ln{s\over u}}
\nonumber\\
&&\qquad\times\ \left\{
i\,f^{aa'c}\,i\,f^{bb'c}\,-\,d^{aa'c}\,d^{bb'c}\,-\,{2\over N_c}
\left(\delta^{ab}\,
\delta^{a'b'}\,+\,\delta^{ab'}\,\delta^{a'b}\,+\,2\delta^{aa'}\,
\delta^{bb'}\right)
\right\}\, ,
\end{eqnarray}
where the $d^{abc}$ are the symmetric $SU(3)$ structure constants and
$\ln(s/u)\simeq -i\pi$ in the physical region.  Both the $m_{4:1}$
subamplitudes (\ref{four}, \ref{five}) and the 
$m_{4:3}$ subamplitudes (\ref{nonp}, \ref{equal})
contribute to this expression.  Note that the term containing
$f^{aa'c}f^{bb'c}$ agrees with the high-energy limit given in 
eq.~(\ref{elasb}).

For quark-quark scattering we find
\begin{eqnarray}
\lefteqn{{\rm Absorp}\, M_4^{1-loop}(B-,A-;A'+,B'+) \ =\  
g^4\,{s\over t} c_{\Gamma}
\left({\mu^2\over -t}\right)^{\epsilon}\,{2\over\epsilon}
\ln{s\over u}}
\nonumber\\
&&\qquad\qquad\times\ \left\{ 
2\,\lambda^c_{a'\bar a}\,\lambda^c_{b'\bar b}\left(N_c+{1\over N_c}\right)
\,-\,
\delta_{a'\bar b}\,
\delta_{b'\bar a}\,\left({N_c^2-1\over N_c}\right)\right\}\, .
\end{eqnarray}
Both the $a_{4:1}$ and $a_{4:2}$ subamplitudes 
(\ref{pippo}, \ref{pippa}) contribute to this expression. 
The absorptive amplitude for quark-antiquark scattering can be obtained
by exchanging $s\leftrightarrow u$.

\vspace{.5cm}

{\sl Acknowledgements}
We should like to thank Lance Dixon, Victor Fadin and Zoltan Kunszt for useful
discussions and Zvi Bern for pointing out to us a misprint in 
eq.~(\ref{pippo}). This work was partly supported by the
NATO Collaborative Research Grant CRG-950176.

\end{document}